\documentclass[conference]{IEEEtran}

\ifCLASSOPTIONcompsoc
  \usepackage[nocompress]{cite}
\else
  \usepackage{cite}
\fi

\usepackage{nicefrac}
\usepackage{siunitx}
\usepackage{array,framed}
\usepackage{booktabs}
\usepackage{tabularx}
\usepackage{
  color,
  float,
  epsfig,
  wrapfig,
  graphics,
  graphicx,
  subcaption
}

\usepackage{textcomp,amssymb}
\usepackage{setspace}
\usepackage{latexsym,fancyhdr,url}
\usepackage{enumerate}
\usepackage{enumitem}
\usepackage[most]{tcolorbox}

\usepackage{algorithm}
\usepackage[noend]{algpseudocode}
\usepackage[symbol]{footmisc}

\usepackage[numbers,sort]{natbib}
\usepackage{tikz}

\usepackage{graphics}
\usepackage{xparse} 
\usepackage{xspace}
\usepackage{multirow}
\usepackage{csvsimple}
\usepackage{balance}
\usepackage{pifont}
\usepackage{svg}

\usepackage{
  tikz,
  pgfplots,
  pgfplotstable
}
\usepackage{xcolor}
\definecolor{darkgreen}{rgb}{0.0, 0.5, 0.0} 
\usepackage[hidelinks]{hyperref}
 \hypersetup{
     colorlinks=true,
     citecolor=darkgreen,     
     linkcolor=red!70!black,       
     urlcolor=blue!70!black       
 }
\usepackage[capitalise,nameinlink, noabbrev]{cleveref}

\newcommand{\email}[1]{{\rm\textsf{\href{mailto:#1}{#1}}}}

\definecolor{darkviolet}{HTML}{9400D3}

\usetikzlibrary{
  shapes.geometric,
  arrows,
  external,
  pgfplots.groupplots,
  matrix
}

\pgfplotsset{compat=1.9}

\usepackage{mathtools}

\DeclareMathAlphabet{\mathcal}{OMS}{cmsy}{m}{n}

\usepackage{svg}

\DeclareGraphicsExtensions{%
    .png,.PNG,%
    .pdf,.PDF,%
    .jpg,.mps,.jpeg,.jbig2,.jb2,.JPG,.JPEG,.JBIG2,.JB2}

\usepackage{xparse}
\newcommand{\bnm}{\begin{newmath}}
\newcommand{\enm}{\end{newmath}}

\newcommand{\bea}{\begin{eqnarray*}}%
\newcommand{\eea}{\end{eqnarray*}}%

\newcommand{\bne}{\begin{newequation}}
\newcommand{\ene}{\end{newequation}}

\newcommand{\bal}{\begin{newalign}}
\newcommand{\eal}{\end{newalign}}

\newenvironment{newalign}{\begin{align}%
\setlength{\abovedisplayskip}{4pt}%
\setlength{\belowdisplayskip}{4pt}%
\setlength{\abovedisplayshortskip}{6pt}%
\setlength{\belowdisplayshortskip}{6pt} }{\end{align}}

\newenvironment{newmath}{\begin{displaymath}%
\setlength{\abovedisplayskip}{4pt}%
\setlength{\belowdisplayskip}{4pt}%
\setlength{\abovedisplayshortskip}{6pt}%
\setlength{\belowdisplayshortskip}{6pt} }{\end{displaymath}}

\newenvironment{newequation}{\begin{equation}%
\setlength{\abovedisplayskip}{4pt}%
\setlength{\belowdisplayskip}{4pt}%
\setlength{\abovedisplayshortskip}{6pt}%
\setlength{\belowdisplayshortskip}{6pt} }{\end{equation}}

\newcounter{ctr}

\newcounter{mytable}
\def\mytable{\begin{centering}\refstepcounter{mytable}}
\def\endmytable{\end{centering}}

\newcounter{myfig}
\def\myfig{\begin{centering}\refstepcounter{myfig}}
\def\endmyfig{\end{centering}}

\newlength{\saveparindent}
\newlength{\saveparskip}
\newcommand{\E}{{\rm I\kern-.3em E}}

\renewcommand{\eqref}[1]{\mbox{Equation~(\ref{#1})}}

\newcommand{\mV}{m\kern-1pt V\xspace}
\newcommand{\Vpp}{V\kern-1pt \textsubscript{pp}\xspace}

\def \part {part}

\renewcommand{\paragraph}[1]{\vspace*{6pt}\noindent\textbf{#1}\;}

\def \blackslug{\hbox{\hskip 1pt \vrule width 4pt height 8pt
    depth 1.5pt \hskip 1pt}}
\def \qed{\quad\blackslug\lower 8.5pt\null\par}

\newcounter{mynote}[section]

\newcommand\ignore[1]{}

\newcounter{rcnote}[section]

\newcounter{mrnote}[section]

\newcounter{fknote}[section]

\newcounter{anote}[section]

\DeclareMathSymbol{\mlq}{\mathord}{operators}{``}
\DeclareMathSymbol{\mrq}{\mathord}{operators}{`'}

\newcommand{\rhf}[2]{R_{f, \gamma}}

\DeclareDocumentCommand{\edist}{o o}{
  \ensuremath{
    \IfNoValueTF{#1}{{d}}{{\sf d}(#1,#2)}
  }
}

\newcommand{\olrk}[1]{\ifx\nursymbol#1\else\!\!\mskip4.5mu plus 0.5mu\left(\mskip0.5mu plus0.5mu #1\mskip1.5mu plus0.5mu \right)\fi}

\NewDocumentCommand{\indseq}{ O{1} O{r} }{{#1}\ldots {#2}}

\makeatletter
\newcommand{\linebreakand}{%
  \end{@IEEEauthorhalign}
  \hfill\mbox{}\par
  \mbox{}\hfill\begin{@IEEEauthorhalign}
}
\makeatother

\definecolor{byzantium}{rgb}{0.44, 0.16, 0.39}

\newif\ifshowcomments
\showcommentstrue 

\begin{document}

\IEEEoverridecommandlockouts

\IEEEpubid{\makebox[\columnwidth]{Accepted to special session track.\hfill} \hspace{\columnsep}\makebox[\columnwidth]{ }}

\fancyhead{}
\def\thetitle{NonSense: On the (In)security of FPGA Anti-tamper Sensors} 




\title{\thetitle}


\author{\IEEEauthorblockN{Mehmet Ali Cetin}
\IEEEauthorblockA{Worcester Polytechnic Institute\\
\small\email{mcetin@wpi.edu}}
\and
\IEEEauthorblockN{Fatemeh Khojasteh Dana}
\IEEEauthorblockA{Worcester Polytechnic Institute\\
\small\email{fdana@wpi.edu}}
\and
\IEEEauthorblockN{Saleh Khalaj Monfared}
\IEEEauthorblockA{Worcester Polytechnic Institute\\
\small\email{skmonfared@wpi.edu}}
\linebreakand
\IEEEauthorblockN{Nima Seyed Aghazadeh}
\IEEEauthorblockA{Worcester Polytechnic Institute\\
\small\email{naghazadeh@wpi.edu}}
\and
\IEEEauthorblockN{Maryam Saadat Safa}
\IEEEauthorblockA{Worcester Polytechnic Institute\\
\small\email{msafa@wpi.edu}}
\and
\IEEEauthorblockN{Shahin Tajik}
\IEEEauthorblockA{Worcester Polytechnic Institute\\
\small\email{stajik@wpi.edu}}}

\date{}

\maketitle

\IEEEpeerreviewmaketitle

\begin{abstract}

FPGAs are widely used in mission-critical systems, where a successful physical attack can have serious consequences. This makes resilience to such attacks a critical security requirement. To defend against active physical attacks, such as glitching attacks, modern FPGAs rely on on-chip sensors that detect these disturbances and trigger responses, such as key zeroization, before a device is compromised. These sensors are deployed as both hard-IP blocks and soft-IP cores, and their trustworthiness is often taken for granted. In this work, we challenge that assumption and argue that the sensors themselves are a fragile and underexplored attack surface. Drawing on recent results, we show that both the detection and response mechanisms of state-of-the-art sensors can be bypassed without invasive tampering. We further argue that sensor reliability is reduced by factors rarely captured in threat models, including the spatial and frequency-dependent behavior of the system’s power delivery network, which can steer a fault to a target while sparing nearby sensors; environmental conditions such as temperature, which shift detection thresholds; and long-term aging, which drifts sensor calibration and may open silent detection gaps. Finally, we propose a set of countermeasures that harden detection and response mechanisms against these effects.

\end{abstract}

\section{Introduction}\label{sec:intro}
FPGAs and FPGA-based systems-on-chip are widely used in applications that require both computational flexibility and strong security guarantees, including aerospace, defense, and secure computing infrastructure. Active physical attacks can threaten their security by manipulating operating conditions or applying external stimuli to influence circuit behavior. Voltage and clock glitches~\cite{buhren2021one} can corrupt computation or bypass security checks, while hibernation attacks~\cite{mitard2026chypnosis,dana2026chypothermia} can halt execution and preserve sensitive state for subsequent extraction. Backscattered side-channel attacks, including laser logic state imaging (LLSI)~\cite{krachenfels2021real} and impedance analysis~\cite{monfared2023leakyohm}, can further expose internal state by measuring a device's response to an externally applied signal.

To detect these attacks, FPGA-based systems can employ sensors that monitor supply voltage, clock activity, temperature, and other physical quantities. These sensors are implemented either as dedicated hard-IP blocks, such as analog-to-digital converters (ADCs) and phase-locked loops (PLLs), or as soft-IP circuits instantiated in the programmable fabric, such as ring oscillators (ROs), time-to-digital converters (TDCs), and delay-based clock monitors. Upon detecting an anomaly, a sensor triggers a protective response, such as clearing sensitive registers or placing the system in a locked state. The effectiveness of this approach depends on two distinct requirements, namely the sensing mechanism must identify the disturbance, and the response must complete before the adversary achieves the security objective. An asserted alarm alone is therefore insufficient to establish protection.

Recent attacks demonstrate that these requirements can fail under adversarial operating conditions. Chypnosis~\cite{mitard2026chypnosis} showed that rapid undervolting can halt clocked computation and prevent clock-dependent protective responses from completing while sensitive state remains stored. Chypothermia~\cite{dana2026chypothermia} demonstrated that cooling can disrupt clock generation and sensor operation, and that temperature manipulation can undermine a delay-based defense against undervolting. GlitchSnipe~\cite{dana2026glitchsnipe} further showed that voltage-glitch sensitivity varies spatially and with the disturbance frequency, reflecting the behavior of the system's power delivery network (PDN). Consequently, a sensor and the logic it protects need not experience the same disturbance, even when they reside on the same die. Together, these results challenge the assumption that an on-chip sensor remains a trustworthy observer throughout an attack.

Sensor behavior also depends on conditions that evolve during ordinary deployment. Temperature changes alter propagation delays and sensor outputs, while device aging can shift the baseline against which anomalies are detected. For fabric-based sensors, placement and routing introduce additional implementation dependence. These effects matter because sensing circuits, alarm paths, and protected logic share physical resources and operating conditions, yet may respond differently to the same change. A threshold calibrated on a device at room temperature may therefore provide different detection margins after prolonged operation or under a different thermal condition. Establishing whether these changes create exploitable gaps requires evaluating their effect on both detection and response.

In this work, we combine recent attack results with experimental characterization of FPGA sensor behavior to examine these dependencies. We distinguish failures of the sensing stage from failures of the protective response and investigate how accelerated aging and temperature variation affect sensor outputs. Our analysis also considers asynchronous sensing and response as a way to reduce dependence on a clock that an attacker may disrupt. However, asynchronous operation does not eliminate dependence on propagation delay, supply voltage, or physical implementation, making environmental and lifetime characterization necessary.

\noindent\textbf{Our Contributions.}
The primary contribution of this work is consolidating recent results on FPGA anti-tamper mechanisms into an analysis that distinguishes sensor detection, alarm generation, and completion of the protective response. We identify how undervolting, cooling, and spatially varying disturbances can undermine these stages.
Moreover, we characterize sensor behavior under accelerated aging, measuring changes in RO and TDC outputs and the loss of valid XADC temperature readings. We also examine the temperature-dependent spatial variation of distributed TDC sensors. These measurements establish changes in sensor behavior and motivate further evaluation of their effects on detection margins and attack success.
Finally, we derive design recommendations for more resilient protection, including asynchronous sensing and response paths, sensor-health monitoring, bounded environmental compensation, and controlled recalibration. We discuss the remaining dependencies on aging, temperature, placement and routing, and shared power-delivery resources.

\section{Technical Background}\label{sec:backgraound}

\subsection{Active Attacks}
\subsubsection{Fault and Glitching Attacks}
The most prominent type of active physical attack is fault/glitching attacks.
Examples include voltage, clock, temperature, and laser fault attacks.
By injecting energy and inducing faults in FPGAs, attackers may bypass certain states and gain unauthorized access to protected states. 
Consequently, they could potentially obtain cryptographic keys, gain privileged access to a service, or simply carry out a denial-of-service attack.

\subsubsection{Backscattered Side-channel Attacks}
Another type of active physical attack is a backscattered side-channel attack, in which the adversary stimulates the chip using external signals (e.g., near-infrared laser beams~\cite{krachenfels2021real} or microwave radiation~\cite{monfared2023leakyohm}) and measures the modulated reflections to infer the internal circuit state or memory contents. Due to their active nature, backscattered attacks typically yield a higher Signal-to-Noise Ratio (SNR) and may, in some situations, recover secrets from a single trace.

\subsubsection{Hibernation and Freezing Attacks}
Recently, a new type of active attack has been introduced in which the adversary tries to stop the system clock to freeze the state of the chip and recover secrets using backscattered attacks.
In these cases, the attacker puts the chip into hibernation by undervolting it (Chypnosis)~\cite{mitard2026chypnosis} or exposing it to extreme cold temperatures (Chypothermia)~\cite{dana2026chypothermia}.

\begin{figure}[t]
    \centering
    \includegraphics[width=\linewidth]{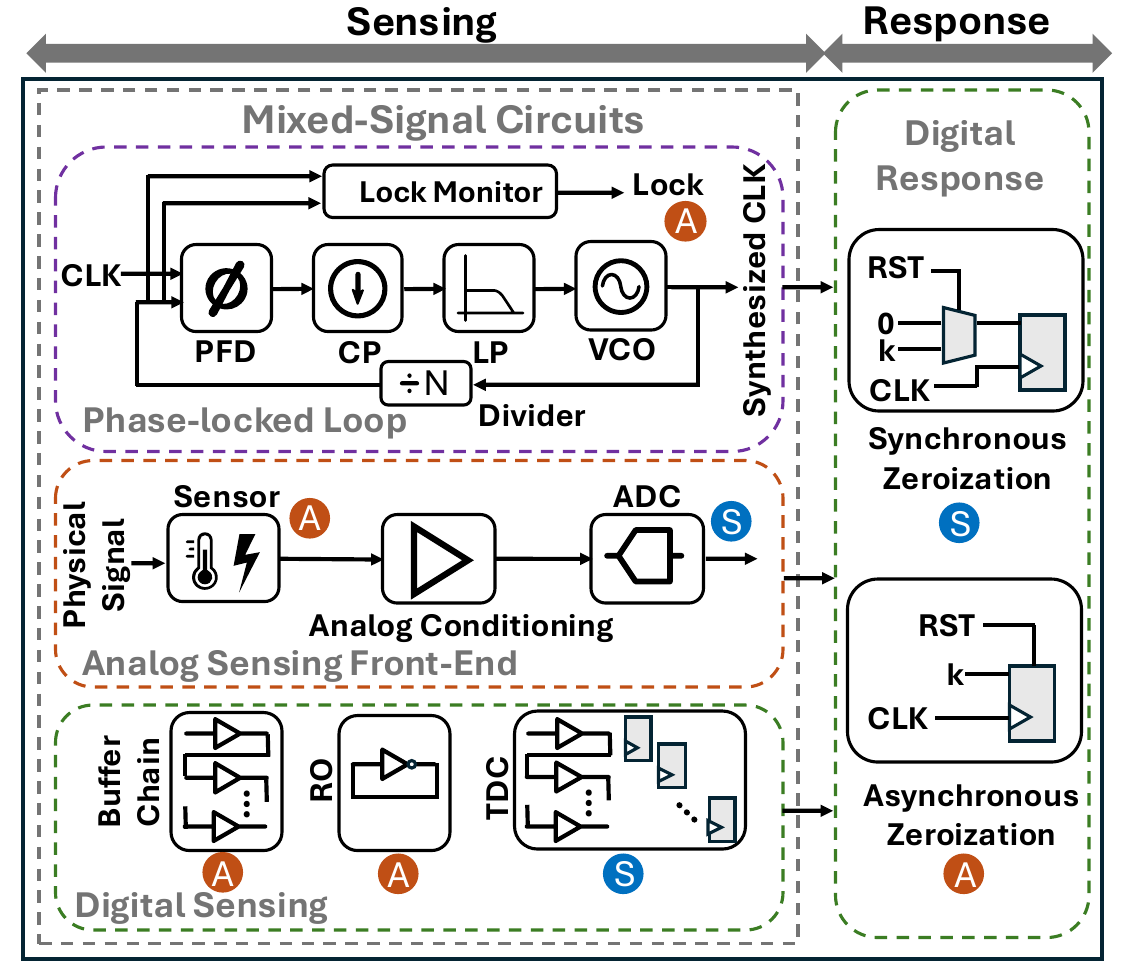}
    \caption{Sensor architecture on modern FPGAs: Sensing and response modules are implemented using mixed-signal and digital circuits, respectively. Both modules can be realized using {\color{blue}{synchronous}} or {\color{orange}{asynchronous}} logic.}
    \label{fig:sensor_module}
\end{figure}

\subsection{Anti-tamper Sensors}

\subsubsection{Sensor Types}
FPGA anti-tamper mechanisms monitor physical quantities or circuit behavior to identify conditions associated with an attack. Voltage sensors detect deviations in supply voltage, while temperature sensors monitor the device's thermal operating conditions. Clock monitors can detect selected anomalies, such as missing edges, prolonged inactivity, or deviations in frequency. Their coverage depends on the monitoring architecture: a PLL lock indicator, for example, reports the status of its associated clock-generation circuitry and should not be assumed to detect every clock disturbance.

Sensors may measure a physical quantity directly or infer it from its effect on circuit behavior. ADC-based monitors digitize signals representing temperature or supply voltage. Fabric-based sensors commonly exploit the dependence of propagation delay on operating conditions, using ring oscillators (ROs)~\cite{gravellier2019high}, time-to-digital converters (TDCs)~\cite{zick2013sensing}, or buffer chains~\cite{dumitru2025borrowed}. These delay-based structures are generally sensitive to multiple influences, including voltage, temperature, and aging. Consequently, a change in their output does not uniquely identify either the physical cause or an attack. Detection requires a decision rule that distinguishes acceptable variation from conditions requiring a protective response.

\subsubsection{Sensor Modules}

Anti-tamper sensors are generally composed of two main components: sensing and response, see Fig.~\ref{fig:sensor_module}. Sensing typically uses mixed-signal circuits that combine analog and digital components to monitor the device's physical state. Examples include PLL-based sensing, which uses the lock signal of a phase-locked loop to detect clock anomalies; an analog sensing front-end, which monitors physical quantities such as voltage and temperature and converts the measurements into digital values using an ADC, and digital sensing, which can use delay-based circuits such as buffer chains, ring oscillators (ROs), or time-to-digital converters (TDCs) to detect anomalies in voltage~\cite{zick2013sensing} and temperature~\cite{lempel2022700}. These sensing components generate an alert when they detect an abnormal condition, which the response part of the sensor module then handles.

Upon receiving the alert signal from the sensing module, the response component performs the predefined action, such as randomizing~\cite{monfared2024laserescape,monfared2024randohm} or zeroizing~\cite{srivastava2019efficient} registers containing sensitive data~\cite{tada2021design}. The response can be either synchronous or asynchronous. In a synchronous response, the circuit waits for a clock edge before clearing the sensitive data. In contrast, an asynchronous response can reset or clear the registers without depending on the system clock. 

\subsubsection{Hard vs. Soft IP Sensors}
Hard-IP sensors are dedicated circuits fabricated at fixed locations in the device. Examples include the XADC in AMD 7~Series devices~\cite {xilinx_xadc}, the System Monitor (SYSMON) in UltraScale and UltraScale+ devices~\cite {amd_sysmon_ug580}, and the Temperature and Voltage Sensor (TVS) in Microchip PolarFire devices~\cite {polarfiremacros,microchip_security}. Their sensing structures and available instances are determined by the device architecture. Depending on the block, users can select measurement channels, configure acquisition and averaging, program alarm thresholds.

Soft-IP sensors, on the other hand, are implemented using programmable resources such as LUTs, carry chains, and flip-flops. An RO converts propagation delay into oscillation frequency, which can be measured over a reference interval. A TDC measures the progress of a transition through a delay line using capture registers; its output depends on both the delay elements and the launch-to-capture timing. A buffer-chain clock monitor detects prolonged inactivity through the settling behavior of delayed signal replicas~\cite{BorrowedTime_github,mitard2026chypnosis,dana2026chypothermia}.

The designer can modify a soft-IP sensor's structure, replication, placement, routing, and decision logic, subject to the available FPGA resources. This flexibility enables distributed monitoring, but also introduces implementation dependence. Sensor characterization must therefore account for the complete implemented measurement path, including clocking and routing, and establish how its baseline and detection margins vary with operating conditions and aging.

\section{Threat Model}
FPGA anti-tamper sensors are designed primarily to thwart a physical adversary.
A physical attacker usually has physical access to the target FPGA and can tamper with the chip's supply voltage, clock, and ambient temperature.
A sophisticated adversary can even go further and expose the FPGA to microwave and laser radiation.
The attacker's goal is either to cause a glitch in the system's operation for privilege escalation or secret recovery, or to mount sophisticated backscattered side-channel attacks to recover the FPGA's internal states.
In cloud FPGA deployments, an attacker might still be able to cause a glitch in the FPGA's operation remotely by uploading a malicious circuit configuration onto FPGAs.
Hence, FPGA sensors should be configured to detect a wide range of attack vectors and respond by wiping secret information or dialing in additional countermeasures.







\section{Sensor Behavior}\label{sec:results}

\subsection{Soft IP sensors}

\begin{figure}[t]
  \centering
  \includegraphics[width=\columnwidth]{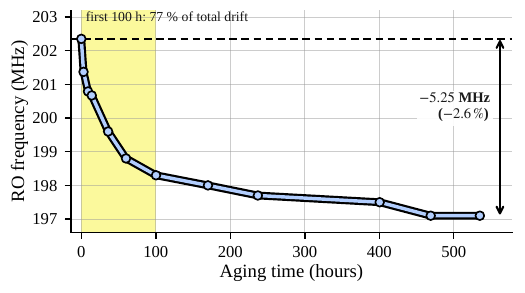}
  \caption{RO frequency of the soft-IP sensor over $535$\,hours of accelerated aging. The frequency drops rapidly at first and then gradually saturates; more than three-quarters of the total drift occurs within the first $100$\,hours.}
  \label{fig:ro-full}
\end{figure}

\begin{figure}[t]
  \centering
  \includegraphics[width=\columnwidth]{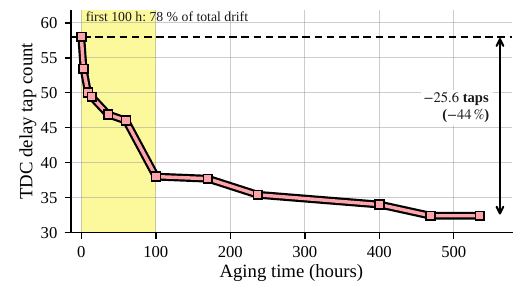}
  \caption{TDC delay tap count over the same aging process. The tap count decreases by $44\,\%$, with a sharp initial drop in the first $100$\,hours followed by a slower decline.}
  \label{fig:tdc-full}
\end{figure}

\subsubsection{Effect of Aging}
To observe how soft-IP sensors drift over the lifetime of a device, we subjected the DUT, an AMD/Xilinx Spartan-7 XC7S50 on a Digilent Arty~S7-50 board, to accelerated aging in the TestEquity TE-107 thermal chamber at $85^{\circ}$C, with the core supply raised from its nominal $1.0$\,V to $1.2$\,V ($20\,\%$ above nominal). In addition, four large ROs distributed across the FPGA fabric were activated, so that the device operated under a heavy workload and a large portion of the fabric was stressed. We characterized two soft-IP sensors, an RO and a TDC, across twelve aging stages spanning $0$--$535$\,hours. After every stress interval, the device was allowed to rest for about one hour to stabilize, and all sensor readings were taken inside the chamber at $25^{\circ}$C to reduce environmental variability. 

Figure~\ref{fig:ro-full} shows the RO frequency throughout the aging process. The RO frequency decreases from approximately $202.3$\,MHz on the fresh device to $197$\,MHz after $535$\,hours, a total reduction of ${\sim}2.6\,\%$. Roughly two-thirds of the total drift accumulates within the first $60$\,hours and more than three-quarters within the first $100$\,hours; the curve then gradually saturates, and the final ${\sim}300$\,hours contribute only about $0.3\,\%$. This shape, an initial rapid drop followed by a gradual saturation, is characteristic of negative bias temperature instability (NBTI)~\cite{wang2011unified,lorenz2012aging}. Under prolonged electrical and thermal stress, interface traps and oxide charges accumulate at the gate oxide, increasing the threshold voltage and reducing transistor drive strength, which slows switching and lowers the RO frequency. For an anti-tamper sensor whose detection threshold is calibrated on a fresh device, this means the largest baseline shift occurs early in the aging process.

\begin{figure}[t]
    \centering
    \includegraphics[width=\linewidth]{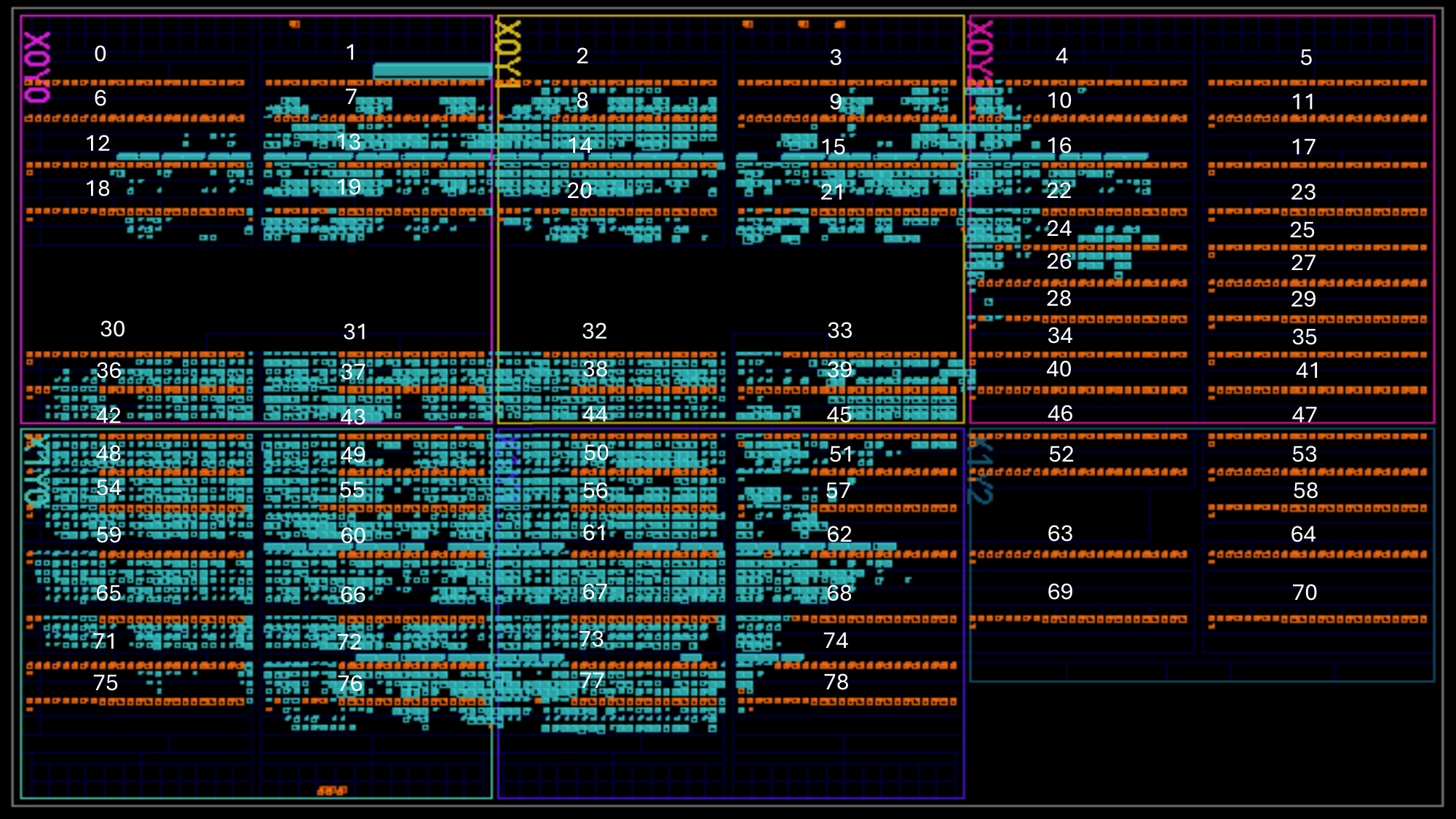}
    \caption{Physical layout of the XC7A35T FPGA, with the locations of the 79 TDC sensors highlighted in orange. 
   }
    \label{fig:tdc_placement}
\end{figure}


The TDC shows the same downward trend, but the relative change in its output is far larger, and its initial degradation is much sharper. The sensor uses two phase-locked clocks that a single on-chip clock manager (MMCM) derives from the board's $100$\,MHz oscillator. A $100$\,MHz clock is launched into a $128$-tap carry chain, built from $32$ cascaded \texttt{CARRY4} elements, and a $300$\,MHz clock samples the state of the chain with the $128$ capture flip-flops. As shown in Figure~\ref{fig:tdc-full}, the delay tap count falls from $58.0$ taps on the fresh device to $32.4$ taps after $535$\,hours, a $44\,\%$ reduction, with $78\,\%$ of the change occurring in the first $100$\,hours, after which the degradation rate slows. As aging slows down signal propagation through the delay line, fewer delay taps are captured within the same measurement window. In other words, the sensor's baseline output, measured under identical conditions at every stage, shifts substantially with age.

\subsubsection{Effect of Temperature}

\begin{figure*}[t]
  \centering
  \begin{subfigure}{0.32\textwidth}
    \includegraphics[angle=-90, width=\linewidth,clip,trim= 0 0 68 20 ,page=9]{figs/glitchsnipe/all_heatmaps_vs_room.pdf}
    \caption{80 ℃}
  \end{subfigure}\hfill
  \begin{subfigure}{0.32\textwidth}
    \includegraphics[angle=-90, width=\linewidth,clip,trim= 0 0 68 20,page=8]{figs/glitchsnipe/all_heatmaps_vs_room.pdf}
    \caption{70 ℃}
  \end{subfigure}\hfill
  \begin{subfigure}{0.32\textwidth}
    \includegraphics[angle=-90, width=\linewidth,clip,trim= 0 0 68 20,page=7]{figs/glitchsnipe/all_heatmaps_vs_room.pdf}
    \caption{60 ℃}
  \end{subfigure}
  \begin{subfigure}{0.32\textwidth}
    \includegraphics[angle=-90, width=\linewidth,clip,trim= 0 0 68 20,page=6]{figs/glitchsnipe/all_heatmaps_vs_room.pdf}
    \caption{50 ℃}
  \end{subfigure} \hfill
  \begin{subfigure}{0.32\textwidth}
    \includegraphics[angle=-90, width=\linewidth,clip,trim= 0 0 68 20,page=5]{figs/glitchsnipe/all_heatmaps_vs_room.pdf}
    \caption{40 ℃}
  \end{subfigure}\hfill
  \begin{subfigure}{0.32\textwidth}
    \includegraphics[angle=-90, width=\linewidth,clip,trim= 0 0 68 20,page=4]{figs/glitchsnipe/all_heatmaps_vs_room.pdf}
    \caption{30 ℃}
  \end{subfigure}
  \begin{subfigure}{0.32\textwidth}
    \includegraphics[angle=-90, width=\linewidth,clip,trim= 0 0 68 20,page=3]{figs/glitchsnipe/all_heatmaps_vs_room.pdf}
    \caption{20 ℃}
  \end{subfigure}\hfill
  \begin{subfigure}{0.32\textwidth}
    \includegraphics[angle=-90, width=\linewidth,clip,trim= 0 0 68 20,page=2]{figs/glitchsnipe/all_heatmaps_vs_room.pdf}
    \caption{10 ℃}
  \end{subfigure}\hfill
  \begin{subfigure}{0.32\textwidth}
    \includegraphics[angle=-90, width=\linewidth,clip,trim= 0 0 68 20,page=1]{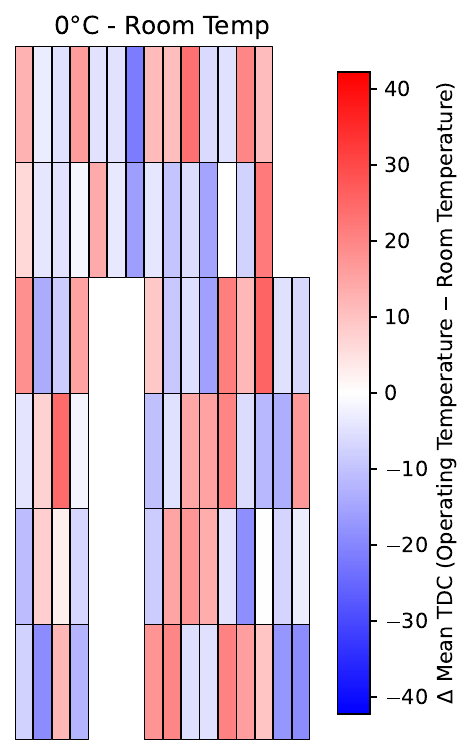}
    \caption{0 ℃}
  \end{subfigure}
   \begin{subfigure}{0.4\textwidth}
    \includegraphics[angle=-90, width=\linewidth,clip,trim= 160 0 20 20,page=1]{figs/glitchsnipe/all_heatmaps_vs_room.pdf}
  \end{subfigure} 
  \caption{Temperature-dependent changes in mean TDC output across 79 sensor instances. Each cell corresponds to a sensor location in Figure~\ref{fig:tdc_placement} and shows
$\Delta\overline{N}(T)=\overline{N}(T)
-\overline{N}(T_{\mathrm{ref}})$, where $\overline{N}$ is the mean TDC tap count and
$T_{\mathrm{ref}}\approx20^\circ$C. Red and blue indicate increases and decreases relative to
the reference measurement; white indicates approximately zero change. Gaps indicate locations without a sensor. All panels use the same color scale.}
  \label{fig:all_heatmaps_vs_room}
\end{figure*}

To evaluate the behavior of soft IP-based sensors under varying environmental conditions, we implemented and characterized a set of TDC sensors across the FPGA fabric. 
In particular, we investigate how temperature variations affect their measurements and spatial behavior across the device.
Similar to the setup in GlitchSnipe~\cite{dana2026glitchsnipe}, we implemented 79 64-bit TDC sensors across the XC7A35T-FTG256 FPGA. 
We placed the TDCs in a 6-row by 16-column grid, although we could not use some locations because the required FPGA resources were unavailable. 
The TDCs were driven by a 250~MHz clock, while the main system clock ran at 100~MHz. We collected the TDC measurements through UART using the ESP-Prog board~\cite{esp_prog_guide}. 
To examine the effect of temperature on the chip's behavior, we placed the NewAE CW305 FPGA board inside the thermal chamber.
To study the effect of temperature on the TDCs, we first performed an experiment at room temperature. 
We calibrated the TDCs at room temperature so that about half of the output bits were ones and the other half were zeros.
Then, we measured the TDCs' behavior over time while changing the temperature from 0°C to 80°C. We compared these results with the room temperature measurements. 
We chose this range because we tested a commercial-grade IC that operates from 0°C to 85°C~\cite{xilinx_ds181}.
Fig.~\ref{fig:all_heatmaps_vs_room} illustrates the behavior of each TDC as a function of temperature, with room temperature corresponding to approximately 20°C. 
A clear Inverse Temperature Dependence (ITD) effect is evident in some locations, leading to faster TDCs as temperature increases.
Similarly, when the temperature decreases, not all locations become faster, as might normally be expected. 
Some locations instead become slower, further demonstrating ITD across the device.

\subsubsection{Effect of Placement and Routing}
Routing and placement can significantly affect FPGA-based sensors, particularly delay-based sensors such as TDCs and ROs. For example, as shown in Fig.~\ref{fig:all_heatmaps_vs_room}, even at a fixed temperature, such as $80^\circ$C, the behavior of the TDCs is not spatially uniform across the FPGA. Some sensors operate faster, while others are slower. 
This spatial variation indicates that the TDCs' response is influenced not only by temperature but also by their physical placement and routing within the device.
Therefore, placement and routing effects should be carefully considered.


\subsection{Hard IP sensors}


\subsubsection{Effect of Aging}

We recorded the XADC temperature output during the accelerated aging experiment described earlier.  We took all measurements after the recovery interval at a chamber temperature of 25$^\circ$C.
As shown in Figure~6, the reported temperature increased from 30.5$^\circ$C on the fresh device to 32.2$^\circ$C after 3~hours, 33.75$^\circ$C after 9~hours, and 36.5$^\circ$C after 14~hours. This represents a 6.0$^\circ$C increase in the reported value. However, constant chamber temperature does not establish constant die temperature. Without an independent thermal reference, the measurements cannot distinguish sensor drift from changes in self-heating.

At the next measurement stage, after 36~hours of stress, the temperature register returned code \texttt{0x000}. 
This code corresponds to $-273.15^\circ$C and therefore indicates an invalid reading rather than a physical die temperature. Zero readback persisted at the subsequent measurement stages through 237~hours, the last recorded XADC
measurement. Meanwhile, the RO and TDC remained measurable through 535~hours.

These observations establish a loss of usable XADC temperature readings under the applied stress conditions, but do not identify the underlying failure mechanism. Continued soft-IP operation does not rule out an XADC-specific conversion,
clocking, reference, or readout failure. No alarm was observed;
however, a low reported code need not trigger a high-temperature alarm, so this observation alone does not establish failure of the alarm circuitry. The result motivates explicit validity
checks beyond threshold alarms and further investigation of the sensor's failure and recovery behavior.

\subsubsection{Effect of Temperature}
Chypothermia~\cite{dana2026chypothermia} investigated hard-IP sensor behavior under liquid-nitrogen cooling on three platforms: the SYSMON of an AMD Zynq UltraScale+ MPSoC, the TVS of a Microchip PolarFire SoC, and the XADC of an AMD Spartan-7 FPGA. The reported experiments illustrate distinct failures of sensing and
clock-monitoring infrastructure under extreme cooling.

\begin{figure}[t]
  \centering
  \includegraphics[width=\columnwidth]{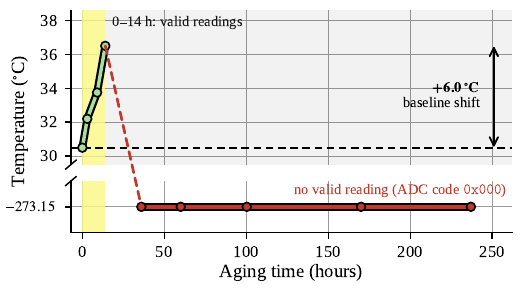}
  \caption{XADC temperature readings during accelerated aging. The reported temperature increases over the first four measurement stages, followed by invalid zero readback from 36~hours through the final XADC measurement at 237~hours. The decoded value of $-273.15^\circ$C represents an invalid code, not a physical temperature.}
  \label{fig:xadc-full}
\end{figure}

On the Zynq UltraScale+ platform, the reported temperature and voltage measurements stopped updating after an internal clock stopped, even though cooling continued. This observation
highlights that plausible sensor readings do not establish continued measurement availability. On the PolarFire SoC, the TVS returned invalid readings while clock activity remained observable during part of the cooling process. The reported
high-temperature alarm also remained inactive despite invalid temperature readings. On the Spartan-7 FPGA, the PLL output clock stopped without the lock signal indicating the failure,
and the XADC malfunctioned.
These results show that hard-IP monitoring and its supporting infrastructure can become unreliable under cooling while sensitive state remains retained.

\subsubsection{Effect of Undervolting}
Chypnosis~~\cite{mitard2026chypnosis} demonstrates that rapid undervolting can interrupt clock generation and prevent protective responses
from completing while sensitive state remains stored in volatile memory. On the evaluated AMD/Xilinx 7~Series and Microchip PolarFire SoC platforms, sufficiently fast supply reductions create conditions in which normal clocked operation
ceases before sensitive state is erased.
Two failure modes must be distinguished: a voltage or clock monitor may fail to signal the disturbance in time, or the disturbance may disable the clock-dependent response after
detection. Thus, an asserted alarm does not establish that protection has completed. These results show that hard-IP monitoring must be evaluated together with its clocking and
response circuitry under the relevant undervolting conditions.

\section{Countermeasures}\label{sec:counter}

Prior work motivates asynchronous sensing and response as a defense against attacks that interrupt clock activity.
Chypnosis~\cite{mitard2026chypnosis} combines delay-based clock-stop detection with asynchronous clearing of complementary registers, allowing the response to complete under the evaluated undervolting conditions. However, Chypothermia~\cite{dana2026chypothermia} shows that
temperature manipulation can undermine the detector's timing.
Local RO-based heating mitigates this effect in the evaluated implementation, at the cost of additional resources and power.
Our measurements motivate three further design considerations.
First, detection thresholds and response timing require margins for temperature variation and aging, with bounded recalibration under trusted conditions. Second, independent health checks should identify invalid readings and stalled conversions rather
than rely exclusively on a sensor's own alarm. Third, placement and routing should be characterized jointly with the protected logic. In other words, spatially distributed sensors may improve coverage, but proximity alone does not ensure equal exposure to a disturbance~\cite{dana2026glitchsnipe}. These recommendations require validation of the complete sensing-to-clearing path under combined attacks and
environmental variation.

\section{Discussion}\label{sec:discussion}
The results examined in this work indicate that the security of FPGA anti-tamper mechanisms depends on the complete path from sensing a disturbance to wiping the sensitive state. A sensor may correctly detect an attack yet provide no protection if its response requires clock edges that the attack suppresses. Conversely, an asynchronous clearing mechanism is ineffective if the sensing or alarm-generation stage stops functioning before asserting the alarm. 
These observations motivate asynchronous sensing and response as a potentially more resilient architecture for attacks that disrupt clock availability. 
A delay-based clock-stop detector can assert an alarm without waiting for another edge of the monitored clock, while asynchronous reset or preset paths can overwrite appropriately implemented sensitive registers after that clock stops. 
The results reported in Chypnosis~\cite{mitard2026chypnosis} support this advantage under the evaluated undervolting conditions. However, an autonomous sensing element, such as an RO, does not make the complete sensor asynchronous if measurement, threshold comparison, or alarm propagation still depends on a vulnerable clock.

Removing clock dependence does not remove dependence on circuit delay, supply voltage, or temperature. 
An asynchronous protection path must still generate and propagate a valid alarm and complete the response before the attacker achieves the security objective or the disturbance disables the remaining response circuitry. 
Consequently, its security must be evaluated using the worst-case latency of the complete sensing-to-clearing path, rather than nominal sensor latency alone. 
A clock-stop detector also does not necessarily detect a short glitch that corrupts computation while the clock continues to run. 
Thus, the benefit of asynchronous operation is limited to the failure modes it addresses and does not protect against all active attacks.

Our aging measurements illustrate why timing margins cannot be assumed constant over the device lifetime. These observations establish substantial changes in sensor baselines, although they do not directly quantify aging in an asynchronous protection path. Depending on the detector and its decision rule, such changes could increase detection latency, cause false alarms, or reduce sensitivity to an attack. 
Moreover, these results support lifetime-aware characterization and recalibration. 
Calibration updates should occur under trusted conditions and remain bounded so that an adversary cannot gradually shift the accepted operating range.

Temperature and physical implementation impose additional constraints. The spatially varying TDC responses mean that a single temperature correction cannot be assumed adequate for every sensor instance. Furthermore, Chypothermia~\cite{dana2026chypothermia} demonstrates that temperature manipulation can undermine a delay-based defense against undervolting, emphasizing the need to evaluate combined disturbances. 
Furthermore, placement and routing affect the sensor's propagation paths and the delivery of its alarm to protected registers. The spatial and frequency-dependent behavior reported in GlitchSnipe~\cite{dana2026glitchsnipe} further indicates that a nearby sensor need not experience the same supply disturbance as the protected logic. 
Sensor placement should therefore be evaluated together with routing, local attack sensitivity, and response latency. Distributing sensors may improve coverage, but replication alone does not eliminate failures caused by shared power, clocking, or alarm circuitry.

\section{Conclusion}\label{sec:conclusion}

This work examined the reliability of hard-IP and soft-IP FPGA anti-tamper sensors by combining recent attack results with measurements of accelerated aging and temperature-dependent spatial variation. The measurements reveal substantial drift in delay-based sensor outputs and a loss of valid ADC readings under accelerated stress. Together with prior demonstrations of clock, sensor, and response failures under undervolting and cooling, these findings show why sensor availability and timely response must be treated as explicit security requirements.
Asynchronous sensing and response can strengthen protection against clock-disruption attacks by allowing alarm generation and sensitive-state clearing to proceed without further edges of the monitored clock. Nevertheless, asynchronous circuits remain subject to aging, temperature, placement and routing, and shared power-delivery effects. Robust protection therefore requires characterization of the complete sensing-to-clearing path, conservative timing and detection margins, sensor-health monitoring, and controlled recalibration. Future evaluations should connect the observed sensor drift to attack success and response completion across devices and combined environmental conditions. FPGA anti-tamper protection should ultimately be assessed by whether it removes sensitive state in time throughout its intended operating lifetime, rather than by whether a sensor raises an alarm under nominal conditions.
We hope this work stimulates interest in more research in security assessment of anti-tamper sensors.

\section*{Acknowledgment}
This work is supported by NSF Grant CNS-2338069.


\bibliographystyle{IEEEtranSN}
\bibliography{ref_fixed_compt}




\end{document}